  \documentstyle[12pt]{article}
  
\catcode`@=11
\def\secteqno{\@addtoreset{equation}{section}%
\def\theequation{\thesection.\arabic{equation}}}
\catcode`@=12
\secteqno
\newcommand{\be}{\begin{equation}}
\newcommand{\ee}{\end{equation}}
\newcommand{\bea}{\begin{eqnarray}}
\newcommand{\eea}{\end{eqnarray}}
\newcommand{\bref}[1]{(\ref{#1})}
\newcommand{\ep}{\epsilon} 
\newcommand{\vf}{\varphi}

\newcommand{\G}{\Gamma} 

           \newcommand{\s}{\sigma}

\def\pa{\partial}

\newcommand{\nn}{\nonumber}

\begin{document}
%\tableofcontents
          \hfill NIIG-DP-01-03

	  \hfill September, 2001

	  \hfill hep-th/??????

%%%%%%%%%%%%%%%%%%%%%%%%%%%%%%%%%%%%%%%%%%%%%%%%%%%%%%%%%%%%%%%
\vskip 20mm

\begin{center} 
{\bf \Large Regularized Quantum Master Equation\\ in the Wilsonian
Renormalization Group}

\vskip 10mm
%\author 
{\large Yuji\ Igarashi, Katsumi\ Itoh and Hiroto\ So$^a$}\par
%}\address{

\medskip
{\it 
Faculty of Education, Niigata University, Niigata 950-2181, Japan\\
$^a$ Department of Physics, Niigata University, Niigata 950-2181, Japan\\
}

\medskip
\date{\today}
\end{center}
%\maketitle
\vskip 10mm
%%%%%%%%%%%%%%%%%%%%%%%%%%%%%%%%%%%%%%%%%%%%%%%%%%%%%%%%%%%%%%%%%%%%%%
\begin{abstract}

Using the Pauli-Villars regularization, we make a perturbative analysis
 of the quantum master equation (QME), $\Sigma =0$, for the Wilsonian
 effective action. It is found that the QME for the UV action determines
 whether exact gauge symmetry is realized along the renormalization
 group (RG) flow. The basic task of solving the QME can be reduced to
 compute the Troost-van Niuwenhuizen-Van Proyen jacobian factor for the
 classical UV action. When the QME cannot be satisfied, the non-vanishing
 $\Sigma$ is proportional to a BRS anomaly, which is shown to be
 preserved along the RG flow. To see how the UV action fulfills the QME
 in anomaly free theory, we calculate the jacobian factor for a pure
 Yang-Mills theory in four dimensions.

\end{abstract}
%%%%%%%%%%%%%%%%%%%%%%%%%%%%%%%%%%%%%%%%%%%%%%%%%%%%%%%%%%%%%%%%%%%%%%
%\pacs{}

\newpage
\setcounter{page}{1}
\setcounter{footnote}{0}
\parskip=7pt
%%%%%%%%%%%%%%%%%%%%%%%%%%%%%%%%%%%%%%%%%%%%%%%%%%%%%%%%%%%%%%%%%%%%%
%%%%%%%%%%%%%%%%%%%%%%%%%%%%%%%%%%%%%%%%%%%%%%%%%%%%%%%%%%%%%%%%%%%%%
%%%%%%%%%%%%%%%%%%%%%%%%%%%%%%%%%%%%%%%%%%%%%%%%%%%%%%%%%%%%%%%%%%%%%
\section{Introduction}

Even if a regularization in field theory is not compatible with a given
symmetry, it does not mean that the symmetry is lost.  The Wilsonian RG
\cite{WilsonKogut} provides us with such an example. Since the
approach\footnote{For recent progress in this subject, see, for example,
ref. \cite{Roma}.}  introduces IR cutoff $k$ to yield the effective
action for lower frequency modes, the standard form of gauge symmetry is
obviously incompatible with the regularization. Nevertheless, as we have
shown in previous papers \cite{Igarashi0,Igarashi1}, an effective but
exact symmetry can be realized along the RG flow.  The key concept for
ensuring the presence of the {\it renormalized symmetry} is the quantum
master equation (QME) in the Batalin-Vilkovisky antifield formalism
\cite{Batalin}. The QME for the Wilsonian action of the IR (macroscopic)
fields~ $\Sigma_{k}[\Phi,\Phi^*]=0$ forms a hypersurface in the theory
space, ie, the space spanned with coupling constants.  An interesting
observation is that, once a theory is found on the hypersurface at some
IR cutoff, it stays on the hypersurface when we lower the cutoff.  In
other words, when we have a UV (microscopic) theory satisfying the QME,
the QME for the IR theory follows and we have the renormalized symmetry.

In our previous works, we assumed (for anomaly free theory) the
existence of a UV action which obeys the QME, $\Sigma[\phi,\phi^*] =0$.
It is of course a non-trivial assumption that the QME holds for a UV
action. The purpose of this paper is to discuss and justify this
assumption within a perturbative framework. In doing this, a UV
regularization should be specified.  In ref. \cite{Igarashi1}, the UV
and IR regularizations were incorporated in a single regulator. Instead,
here we treat two regularizations independently and use the
Pauli-Villars (PV) scheme for the UV regularization.  This allows us to
extract the dependence on the UV regularization out of
$\Sigma_{k}[\Phi,\Phi^*]$. Actually, we show that, at the one-loop
level, $\Sigma_{k}[\Phi,\Phi^*]$ for the Wilsonian action called the
average action becomes $\Sigma[\phi_0,\phi^*]$ for some classical field
configuration $\phi_0$.  The latter Ward-Takahashi (WT) operator
$\Sigma[\phi_0,\phi^*]$ is solely determined by the UV theory and it is,
as a functional, independent of the IR cutoff.  The IR cutoff comes in
only through the classical configuration.  Since we may confirm
$\Sigma[\phi_0,\phi^*]=0$ in an anomaly free theory, we conclude that
$\Sigma_{k}[\Phi,\Phi^*]=0$ for {\it any} IR cutoff $k$.  This
demonstrates the presence of the renormalized symmetry along the RG
flow.

The UV regularized WT operator $\Sigma[\phi,\phi^*]$ naturally arises
when the PV fields are integrated out.  For a given classical UV action
with the standard BRS symmetry, we may look for a quantum UV action,
which solves the QME.  The procedure itself is quite straightforward.
Our regularized expression for $\Sigma[\phi,\phi^*]$ contains the
regularized jacobian factor given by Troost-van Niuwenhuizen-van Proyen
(TVV) \cite{TVV} originally for the calculation of anomalies in the
antifield formalism. Since the PV mass terms breaks the standard BRS
symmetry, this jacobian factor generates possible ``anomaly'' terms.  In
the absence of cohomologically nontrivial anomaly, such ``anomaly''
terms are all superficial, and should be written as BRS transformation
of some local counter terms. Once we find the counter terms, we may
satisfy $\Sigma[\phi,\phi^*]=0$.  Actually, this task was already done
for a pure Yang-Mills theory in ref. \cite{DeJonghe}.

Our formulation should be compared with the so-called fine-tuning
procedure [8-10]. There, one fixes gauge
non-invariant counter terms to compensate the symmetry breaking terms
generated by the regularization, using the effective Ward-Takahashi or
Slavnov-Taylor identity.  In doing this, the UV and IR regularizations
were introduced with a single regulator, and the effective WT identity
for the Legendre action involving the regulator was analyzed for a fixed
IR cutoff $k$.  We will see that our formulation has advantages to the
fine-tuning on two points.  First is the separation of the IR and UV
regularizations.  Second is the use of the Wilsonian action or the
average action, which makes the expression of the WT operator simpler.

Let us emphasize that our formulation applies even to the case that a
genuine anomaly is present in the UV theory. In this case, the
non-vanishing WT operator $\Sigma[\phi_0,\phi^*]$ itself is proportional
to the anomaly.  Since it is equal to the WT operator
$\Sigma_{k}[\Phi,\Phi^*]$ for the Wilsonian action, it implies that the
BRS anomaly is preserved along the RG flow.

We also stress that our discussion given in this paper, after a slight
modification, is applicable to global symmetries as well.

This paper is organized as follows. In the next section, we briefly
 summarize some results of our formalism needed to perform subsequent
 perturbative computation. In section 3, the PV regularization scheme is
 applied to obtain UV regularized one-loop expressions of the WT
 operator for the Wilsonian as well as the Legendre effective action.
 In section 4, the quantum UV action is constructed for a pure
 Yang-Mills theory, by using the TVV formalism.  The last section is
 devoted to discussion.

%%%%%%%%%%%%%%%%%%%%%%%%%%%%%%%%%%%%%%%%%%%%%%%%%%%%%%%%%%%%%%%%%%%%%
%%%%%%%%%%%%%%%%%%%%%%%%%%%%%%%%%%%%%%%%%%%%%%%%%%%%%%%%%%%%%%%%%%%%%
\section{The average action and the quantum master equation}

Let us consider a gauge theory and its gauge-fixed action in
D-dimensional Euclidean space. All fields including ghosts and those for
gauge fixing are denoted collectively by $\phi^{A}$.  The index $A$
labels Lorentz indices $\mu, \nu$ of tensor fields, the spinor indices
of the fermions, and/or an index distinguishing different types of the
generic fields.  The Grassmann parity for fields is expressed by
$\ep(\phi^{A})=\ep_{A}$.  The antifields $\phi_{A}^{*}$ with the opposite
Grassmann parity $\ep(\phi_{A}^{*})=\ep_{A}+1$ are introduced to make
canonical conjugate pairs
\bea
\left(\phi^{A},~\phi_{B}^{*}\right)_{\phi} = \delta_{~B}^{A}\equiv \delta_{AB} (2\pi)^{D}\delta(p-q),
\label{ab0}
\eea
where the antibracket is defined by 
\bea
\left(F,~G\right)_{\phi} &\equiv&\frac{{\partial}^{r} F}{\partial \phi^{A}} 
\frac{{\partial}^{l} G}{\partial \phi^{*}_{A}}
-\frac{{\partial}^{r} F}{\partial \phi^{*}_{A}} 
\frac{{\partial}^{l} G}{\partial \phi^{A}}\nn\\
&=& \int \frac{d^{D}p}{(2\pi)^D}\left[
\frac{{\partial}^{r} F}{\partial \phi^{A}(-p)} 
\frac{{\partial}^{l} G}{\partial \phi^{*}_{A}(p)}
-\frac{{\partial}^{r} F}{\partial \phi^{*}_{A}(-p)} 
\frac{{\partial}^{l} G}{\partial \phi^{A}(p)}\right].
\label{ab1}
\eea
In this paper we use a matrix notation\footnote{For details of this
notation, see ref. \cite{Igarashi1}.} in which the index $A$ also denotes
momentum.
 
Let $S[\phi,~\phi^*]$ be a gauge-fixed action. The gauge is fixed by the
canonical transformation generated by a gauge fermion.  In this
gauge-fixed basis, the antifields remain intact. Our formalism is based on
a continuum analog of the block-spin transformation \cite{Wetterich},
where $\{\phi_{A},~\phi_{A}^{*}\}$ are identified with the UV
(microscopic) variables. They are transformed into the IR (macroscopic)
variables $\{\Phi_{A},~\Phi_{A}^{*}\}$ by a coarse-graining
procedure. To perform the block-spin transformation, we introduce a test
function $f_{k}(p^2)$ for the coarse-graining and cutoff functions
$R^{k}_{AB}$. They depend on an IR cutoff $k$. As a matrix, $R^{k}_{AB}$
is invertible\footnote{The invertible matrix $R^{k}_{AB}$ has
the signature $\ep(R^{k}_{AB})= \ep_{A} + \ep_{B}$. This matrix
and its inverse satisfy
$R^{k}_{BA}=(-)^{\ep_{A}+\ep_{B}+\ep_{A}\ep_{B}} R^{k}_{AB}$ and
$(R_{k}^{-1})^{BA}= (-)^{\ep_{A}\ep_{B}}(R_{k}^{-1})^{AB}$.} and
it may be chosen as
\bea
(R^{k})_{AB}(p,-q) &=&
({\cal R}^{k})_{AB}(p)(2\pi)^{D} \delta(p-q), \nn\\
({\cal R}^{k})_{AB}(p) &=& \frac{{\bar{\cal R}}_{AB}(p)}
{f_{k}(1 - f_{k})},
\label{cutoff}
\eea
where ${\bar{\cal R}}_{AB}(p)$ are assumed to be polynomials in $p$. The
function $f_{k}$ behaves as $f_{k}(p^2)
\approx 0$ for $k^2 < p^2$, and $f_{k}(p^2) \approx 1$ otherwise. 

We consider a gaussian integral
\bea
1 &=& {N}_{k} \int {\cal D} \Phi {\cal D}
\Phi^{*}\prod_{A}\delta \left(\Phi^{*}_{A}- f_{k}^{-1}
\phi^{*}_{A}\right)~\nn\\
&~&~~~~~~~~~~~\times \exp \Biggl\{-\frac{1}{2 \hbar}
\left(\Phi^{A} - f_{k}\phi^{A}\right)R^{k}_{AB} 
\left(\Phi^{B} - f_{k}\phi^{B} \right)\Biggr\},
\label{gauss}
\eea
with a normalization constant ${N}_{k}$, and rewrite the path integral of
the UV fields,\footnote{For simplicity, we suppress the source terms for
the UV fields which were included in ref. \cite{Igarashi1}.}
\bea
Z &=& \int {\cal D}\phi {\cal D} \phi^{*} \prod_{A}\delta (\phi^{*}_{A})
\exp\left(-S[\phi,~\phi^*]/{\hbar}\right) \nn \\
&=& \int {\cal D} \Phi 
{\cal D}\Phi^{*}\prod_{A}\delta\left(\Phi^{*}_{A}\right) 
\exp \left(-W_{k}[\Phi,~\Phi^*]/{\hbar}\right).
\label{part-func1}
\eea
Here $W_{k}$ is the Wilsonian effective action, called as the average
action \cite{Wetterich}.  The subtracted average action,
\bea
 {\hat W}_{k}[\Phi,~\Phi^{*}] = W_{k}[\Phi,~\Phi^{*}] - 
\frac{1}{2}\Phi^{A} {R}^{k}_{AB}\Phi^{B},
\label{What0}
\eea
is the generating functional of the connected cutoff Green functions of
the UV fields:
\bea
&{}&\exp\left(-{\hat W}_{k}[\Phi,~\Phi^*]/\hbar \right) = N_{k}
\int {\cal D} \phi {\cal D} \phi^{*}\prod_{A}\delta
\left(\Phi^{*}_{A}- f_{k}^{-1}
\phi^{*}_{A}\right)\nn\\
&{}&~~~~~~~~~\times \exp\Bigl\{-(S[\phi,\phi^*] +\frac{1}{2}\phi^{A} f_{k}^{2}R^{k}_{AB}\phi^{B}-\Phi^{B}f_{k} R^{k}_{BA}\phi^{A})/\hbar \Bigr\}.
\label{average-action1}
\eea

In the average action, we find that $\Phi^A(p)\approx
f_{k}(p^2)\phi^A(p)$, ie, the IR fields approximate ``the averaged
fields.''  For the antifields, we impose the relation
$\Phi^{*}_{A}=f_{k}^{-1}\phi^{*}_{A}$.  In \bref{average-action1}, the
terms $\phi^A f_{k}^{2}{R}^{k}_{AB} \phi^B$ act as an IR regulator, and
the regularization is constructed in such a way that the integration of the
UV fields is performed for those modes with momenta larger than $k$.
For each UV field $\phi^{A}$, the combination
\bea
 \Phi^{B}  f_{k} R^{k}_{BA} \equiv j_{A}
\label{source}
\eea
acts as the source.

The Legendre effective action is given by
\bea
{\hat \Gamma}_{k}[\vf,~\vf^{*}] \equiv{\hat W}_{k}
[\Phi,~\Phi^{*}] +j_{A}\vf^{A},
\label{Ghat}  
\eea
where the classical UV fields $\vf^A$ are defined as the expectation
values of the UV fields $\phi^A$ in the presence of the sources
$j_{A}$. The antifields are related each other as 
\bea
\vf_{A}^{*}\equiv \phi_{A}^{*}= f_{k} \Phi_{A}^{*}.
\label{vf*}
\eea
Another Legendre effective action, directly related to the average action,
is given by
\bea
{\Gamma}_{k}[\vf,~\vf^{*}] &\equiv& {\hat \Gamma}_{k}[\vf,~\vf^{*}] -
\frac{1}{2}  \vf^{A} f_{k}^{2}R^{k}_{AB}\vf^{B} \nn\\
 &=& {W}_{k}[\Phi,~\Phi^{*}] -
\frac{1}{2}\left(\Phi^{A}-  f_{k}\vf^{A}\right)  
R^{k}_{AB}\left(\Phi^{B}-f_{k} 
 \vf^{B}\right).
\label{G}
\eea

We now discuss how the renormalized BRS symmetry is realized along the
RG flow.  To this end, we define the WT functional $\Sigma$ for the UV
fields:
\bea
 \Sigma[\phi,~\phi^{*}]\equiv {\hbar}^{2} \exp(S/\hbar)\Delta_{\phi}\exp(-S/\hbar)=
\frac{1}{2}\left(S,~S\right)_{\phi} - \hbar \Delta_{\phi} S,
\label{qme1}
\eea
where the $\Delta$-derivative is given by
\bea
\Delta_{\phi} &\equiv& (-)^{\ep_{A}+1}\frac{\partial^r}{\partial \phi^{A}}
\frac{\partial^r}{\partial \phi^{*}_{A}}= (-)^{\ep_{A}+1}\int \frac{d^{D}p}{(2\pi)^D}\frac{\partial^r}{\partial \phi^{A}(-p)}
\frac{\partial^r}{\partial \phi^{*}_{A}(p)}.
\label{Delta1}
\eea
We also define the WT operator for the IR fields,
\bea
\Sigma_{k}[\Phi,~\Phi^{*}]\equiv {\hbar}^{2}\exp(W_{k}/\hbar) 
\Delta_{\Phi}\exp(-W_{k}/\hbar)= \frac{1}{2}\left(W_{k},~W_{k}\right)_{\Phi} - \hbar
\Delta_{\Phi}W_{k},
\label{Sigma_k}
\eea
where $(~,~)_{\Phi}$ and $\Delta_{\Phi}$ denote the antibracket and the
$\Delta$-derivative for the IR fields.  Then, one obtains \cite{Igarashi1}
\bea
\left< \Sigma[\phi,~\phi^{*}]\right>_{\phi}
&\equiv& {\hbar}^{2}\exp(W_{k}/\hbar)N_{k}\int {\cal D} \phi {\cal D}
\phi^{*}\prod_{A}\delta\left(f_{k}
\Phi^{*}_{A}- \phi^{*}_{A}\right)
\nn\\
&{}& \times 
\exp\Bigl\{-\frac{1}{2\hbar}(\Phi -f_{k}\phi)^{A} R^{k}_{AB}(\Phi -f_{k}\phi)^{B}\Bigr\}
\Delta_{\phi}\exp(-S/\hbar) \nn\\
&=&\Sigma_{k}[\Phi,~\Phi^{*}].
\label{QMEW}
\eea
Therefore, if the UV action satisfies the QME $\Sigma[\phi,~\phi^*]=0$,
 the average action automatically obeys the QME,
 $\Sigma_{k}[\Phi,~\Phi^{*}]=0$ for any $k$.

The WT operator $\Sigma_{k}[\Phi,~\Phi^{*}]$ may be expressed in terms
of the Legendre effective action \cite{Igarashi1},
\bea
\Sigma_{k}[\Phi,~\Phi^{*}]=\frac{\pa^{r}{\G}_{k}}
{\pa \vf^{A}}~\frac{\pa^{l}{\G}_{k}}{\pa
\vf_{A}^{*}} - \hbar~f_{k}^{2}R^{k}_{AC} 
\left(\frac{\pa^{l} \pa^{r} {\G}_{k}}
{\pa \vf_{C}^{*} \pa \vf^{B}}\right)
\Biggl(\left[f_{k}^{2} R^{k}+
\frac{\pa^{l} \pa^{r} {\G}_{k}}{\pa \vf \pa\vf}\right]^{-1}\Biggr)^{BA}
 .
\label{brokenST}
\eea
When applied to a pure Yang-Mills theory, eq. \bref{brokenST} reduces 
to the ``modified Slavnov-Taylor'' identity obtained by Ellwanger \cite{Ellwanger0}. 
In the next section, we derive perturbative expressions of \bref{QMEW} and \bref{brokenST} at one-loop level.

%%%%%%%%%%%%%%%%%%%%%%%%%%%%%%%%%%%%%%%%%%%%%%%%%%%%%%%%%%%%%%%%%%%%%%
%%%%%%%%%%%%%%%%%%%%%%%%%%%%%%%%%%%%%%%%%%%%%%%%%%%%%%%%%%%%%%%%%%%%%%
\section{The Pauli-Villars regularization}

In the previous section, we presented a brief summary of our formalism
with an IR regularization.  For perturbative analysis, a UV
regularization should also be included.  This can be done, as in
previous work \cite{Igarashi1}, by taking the test function $f_{k}$ and
the cutoff functions $R^{k}$ that depend on both IR and UV
cutoffs. Instead, here we use the Pauli-Villars (PV) method discussed by
TVV \cite{TVV}, independently of the IR regularization.  Application of
the method to \bref{QMEW} and \bref{brokenST} allows us to extract the
dependence on the UV regularization of the WT operator $\Sigma_{k}$.
This makes our analysis simpler than previous perturbative studies
\cite{Bonini0,Bonini1}.

%%%%%%%%%%%%%%%%%%%%%%%%%%%%%%%%%%%%%%%%%%%%%%%%%%%%%%%%%%%%%%%%%%%%%%
\subsection{The WT operator for the average action}

Let us make a one-loop evaluation of our relation \bref{QMEW}. 
We begin with a UV action
\bea
S[\phi,\phi^*] = S^{(0)}[\phi,\phi^*] + \hbar S^{(1)}[\phi,\phi^*],
\label{UV}
\eea
where $S^{(0)}$ and $S^{(1)}[\phi,\phi^*]$ denote a gauge-fixed
classical action and its counter action, respectively.  The BRS
invariance of the UV theory is expressed as the classical master
equation,
\bea
\left(S^{(0)},~S^{(0)}\right)_{\phi}=0.
\label{cme}
\eea

{}For each field $\phi^{A}$ entering in a loop, introduced is a PV
partner $\chi^{A}$ which has the same statistics as $\phi^{A}$, but the
path integral is formally defined in such a way that a minus sign is
produced in loops. For the PV fields $\chi^{A}$, their antifields
$\chi_{A}^{*}$ are also introduced. The BRS transformation of the PV
sector is defined such that the total measure is invariant, and the
massless part of the PV action is invariant. See ref. \cite{TVV} for more 
details of the PV scheme. The PV action is given by
\bea
S_{\rm PV}[\chi,\chi^{*},\phi,\phi^{*}] &=& S^{(0)}_{\rm PV}[\chi,\chi^{*},\phi,\phi^{*}] + S_{\rm PV}^{\Lambda}[\chi,\chi^{*},\phi,\phi^{*}],\nn\\
S^{(0)}_{\rm PV}&=& \frac{1}{2}\chi^{A}L_{AB} \chi^{B} 
+ \chi^{*}_{A} K^{A}_{~B}\chi^{B}
+ \frac{1}{2}\chi^{*}_{A} M^{AB}\chi^{*}_{B}, \nn\\
S^{\Lambda}_{\rm PV} &=& \frac{1}{2}\Lambda~\chi^{A} T_{AB} \chi^{B}, 
\label{PV-action-1}
\eea
where $S_{\rm PV}^{(0)}$ is the massless action, and $S^{\Lambda}_{\rm
PV}$ is a mass term with invertible matrix $T_{AB}$ which may depend on
the UV fields $\phi^{A}$ but not on the antifields $\phi_{A}^{*}$.  
We take the mass $\Lambda =M^2$ for bosons and $\Lambda =M$ for fermions. 
The
matrices needed to specify the massless action are given by
\bea
 L_{AB} = \frac{\partial^{l}}{\partial \phi^{A}}
 \frac{\partial^{r} S^{(0)}}{\partial \phi^{B}}, 
~~~~K^{A}_{~B} =
\frac{\partial^{l}}{\partial \phi^{*}_{A}}
 \frac{\partial^{r} S^{(0)}}{\partial \phi^{B}},  
 ~~~~ M^{AB}=  
\frac{\partial^{l}}{\partial \phi^{*}_{A}}
 \frac{\partial^{r} S^{(0)}}{\partial \phi^{*}_{B}}. 
\label{PV-action-2}
\eea

Let $\Sigma[\phi,\phi^{*}, \chi,\chi^{*}]$ be the WT operator for the
total UV action $S[\phi,\phi^{*}] + S_{\rm
PV}[\chi,\chi^{*},\phi,\phi^{*}]$:
\bea
\Sigma[\phi,\phi^{*}, \chi,\chi^{*}] &=& 
\frac{1}{2}\left[(S+S_{\rm PV},~ S+S_{\rm PV})_{\phi} + 
(S+S_{\rm PV},~ S+S_{\rm PV})_{\chi}\right]\nn\\
&{}& - \hbar \left(\Delta_{\phi} +\Delta_{\chi}\right)(S+S_{\rm PV}).
\label{Sigma-phi-chi}
\eea
We may define the UV regularized WT operator by
\bea
\Sigma_{\rm reg}[\phi,\phi^{*}]&\equiv&  \left<\Sigma [\phi,\phi^{*},\chi,\chi^{*}]\right>_{\chi}\nn\\
&\equiv& \frac{\int 
{\cal D} \chi {\cal D} \chi^{*}\prod_{A}\delta
(\chi^{*}_{A})\Sigma [\phi,\phi^{*},\chi,\chi^{*}]
\exp\biggl(-S_{\rm PV}[\chi,\chi^{*},\phi,\phi^{*}]/\hbar \biggr)}{\int 
{\cal D} \chi {\cal D} \chi^{*}\prod_{A}\delta(\chi^{*}_{A})
\exp\biggl(-S_{\rm PV}[\chi,\chi^{*},\phi,\phi^{*}]/\hbar \biggr)}.
\label{reg-Sigma-1}
\eea
At one-loop order, the expression of this WT operator can be simplified
as follows. First, the requirement of a mode-by-mode cancellation
between the jacobian factors for $\phi$ and $\chi$ leads to
\bea
\left(\Delta_{\phi} +\Delta_{\chi}\right)\left(S+ S_{\rm PV}\right)= 0.
\label{cancel-measure}
\eea
Second, it follows from \bref{cme} that
\bea
\left[\frac{1}{2}(S^{(0)}_{\rm PV}, S^{(0)}_{\rm PV})_{\chi}+ 
(S^{(0)}_{\rm PV}, S^{(0)})_{\phi}\right]_{\chi^{*}=0} \propto \chi^{A}
\frac{\pa^{l}}{\partial \phi^{A}}\frac{\pa^{r}}{\partial \phi^{B}}\left( S^{(0)},~S^{(0)}\right)_{\phi}\chi^{B}=0.
\label{massless}
\eea
Third, since $\left<\chi \chi \right>_{\chi} \sim O(\hbar)$, one finds 
$\left<(S^{(0)}_{\rm PV}, S^{(0)}_{\rm PV})_{\phi}\right>_{\chi} \sim
O(\hbar^2)$. Thus, the remaining terms are given by
\bea
\left<\Sigma [\phi,\phi^{*},\chi,\chi^{*}]\right>_{\chi} = \hbar
(S^{(0)}, ~S^{(1)})_{\phi} + \left< (S^{\Lambda}_{\rm PV}, ~S^{(0)})_{\phi} + (S^{\Lambda}_{\rm PV},~S^{(0)}_{\rm PV})_{\chi}
\right>_{\chi},
\label{Sigma-phi-chi-2}
\eea
where
\bea
\left(S^{\Lambda}_{\rm PV},~S^{(0)}\right)_{\phi} &=& \frac{\Lambda}{2}\chi^{A}
(-)^{\epsilon_B} \left(T_{AB}, S^{(0)}\right)_{\phi}\chi^{B},\nn\\ 
%= \frac{1}{2}\chi^{A} (-)^{B} \delta T_{AB}\chi^{B},\nn\\
\left(S^{\Lambda}_{\rm PV},~S^{(0)}_{\rm PV}\right)_{\chi}&=& \Lambda ~\chi^{A} T_{AC} K^{C}_{~B} \chi^{B}.
\label{blacket}
\eea

The integration of the PV fields gives
\bea
\left< (S^{\Lambda}_{\rm PV},S^{(0)})_{\phi} + (S^{\Lambda}_{\rm
PV},S^{(0)}_{\rm PV})_{\chi}\right>_{\chi}
\hspace{-2mm}
=\Lambda\left<\chi^{A} (T{\cal K})_{AB} \chi^{B}\right>_{\chi}
\hspace{-2mm}
=-\hbar \Lambda ~{\rm tr}\left(T{\cal K}[L + \Lambda T]^{-1}\right),
\label{chi-integral}
\eea
where 
\bea
{\cal K}^{A}_{~B} &=& K^{A}_{~B} + \frac{1}{2}(T^{-1})^{AC}(-)^{\epsilon_B} 
\left(T_{CB}, S^{(0)}\right)_{\phi}.
% \delta T_{CB}
\label{cal-K}
\eea
We then obtain
\bea
\Sigma_{\rm reg}[\phi, \phi^{*}] &=& 
\hbar (S^{(0)}, S^{(1)})_{\phi} - \hbar \left(\Delta S^{(0)}\right)_{\rm reg},
\nn\\
\left(\Delta S^{(0)}\right)_{\rm reg} &\equiv&
{\rm tr}
\left({\cal K}[{1+ {\cal O}/\Lambda}]^{-1}\right),~~~~{\cal O} \equiv T^{-1} L.
\label{reg-Sigma-2}
\eea
The WT operator for the IR fields $\Sigma_{k}[\Phi,~\Phi^{*}]$ is given
by the functional average of \bref{reg-Sigma-2} over the UV
fields. Since the $\Sigma_{\rm reg}$ is proportional to $\hbar$, we do
not need to make the $\phi$ integration.  Expanding the UV fields around 
their classical fields $\phi_{0}$ determined by the saddle-point
equations, we obtain
\bea
\Sigma_{k}[\Phi,~\Phi^{*}] = \Sigma_{\rm reg}[\phi_{0}, \phi^{*}].
\label{one-loop-result}
\eea
This relation is our main result and it has important implications to be
discussed in the next subsection.

\subsection{The WT operator for the Legendre action}

In order to compare our formalism with the previous fine-tuning analysis
[8-10], we discuss regularized expression of the WT identity for the
Legendre effective action.  The general relation \bref{brokenST}
suggests that one should obtain the same conclusion as the previous
subsection.  We derive the one-loop version of \bref{brokenST} directly
in order to see how the IR regulator can be separated from the WT
identity. To this end, we consider the partition function for the
regularized version of the subtracted average action,
\bea
{\hat Z}[j,~\phi^*] &=& \exp\left(-{\hat W}_{k\Lambda}[j,~\phi^*]/\hbar \right) 
= N_{k}
\int {\cal D} \phi {\cal D} \chi {\cal D} \chi^{*}\prod_{A}\delta
(\chi^{*}_{A})\nn\\ &{}& ~~~~~~~ \times 
\exp- \biggl(S_{\rm tot}[\phi,\phi^*, \chi,\chi^{*}]-j_{A}\phi^{A}\biggr)/\hbar,
\label{reg-what}
\eea
where the sources $j_{A}$ are related to the IR fields as in
\bref{source}, and the total action is given by
\bea
S_{\rm tot}[\phi,\phi^*, \chi,\chi^{*}] &=& 
S_{k}[\phi,\phi^*] + S_{\rm PV}[\chi,\chi^{*},\phi,\phi^{*}],\nn\\
S_{k}[\phi,\phi^*]&=& S[\phi,\phi^*] + \frac{1}{2}
\phi^{A} f_{k}^{2}R^{k}_{AB}\phi^{B}.
\label{S-tot}
\eea
In this subsection, the regularized average action is expressed as
${\hat W}_{k\Lambda}$ to indicate that both the IR and UV regularizations are
introduced.  The regularized Legendre action, which generates the 1PI
cutoff vertex functions of the UV fields, is given by
\bea
{\hat \Gamma}_{k\Lambda}[\vf,~\vf^{*}] &=&{\hat W}_{k\Lambda}
[j,~\phi^{*}] +j_{A}\vf^{A}.
\label{Ghat2}  
\eea
One obtains the WT identity as is shown in Appendix A: 
\bea
&{}& \frac{1}{2}\left({\hat \Gamma}_{k\Lambda}, {\hat
\Gamma}_{k\Lambda}\right)_{\vf} \nn\\
&{}&~~=
\frac{\int {\cal D} \phi {\cal D} \chi {\cal D} \chi^{*}\prod_{A}\delta
(\chi^{*}_{A})~\Xi_{\rm tot}~\exp \Bigl\{-\left(S_{\rm
tot}-\frac{\partial^{r}{\hat
\Gamma}_{k}}{\partial\vf^{A}}(\phi-\vf)^{A}\right)/\hbar \Bigr\}}
{\int {\cal D} 
\phi {\cal D} \chi {\cal D} \chi^{*}\prod_{A}\delta
(\chi^{*}_{A})\exp\Bigl\{ -\left(S_{\rm tot}-
\frac{\partial^{r}{\hat \Gamma}_{k}}{\partial \vf^{A}}
(\phi-\vf)^{A}\right)/\hbar \Bigr\}} 
\label{Gamma-Gamma}
\eea
with the total WT operator
\bea
\Xi_{\rm tot}[\phi,\phi^*, \chi,\chi^{*}] &\equiv& \frac{1}{2}\left[
(S_{\rm tot}, S_{\rm tot})_{\phi} + (S_{\rm tot},
S_{\rm tot})_{\chi}\right] -\hbar
\left(\Delta_{\phi}+\Delta_{\chi}\right)S_{\rm tot}\nn\\
&=& \frac{1}{2}\left[
(S_{\rm tot}, S_{\rm tot})_{\phi} + (S_{\rm tot},
S_{\rm tot})_{\chi}\right].
\label{Xi}
\eea
Here we used the cancellation condition of the jacobian factors 
\bref{cancel-measure}. In contrast to \bref{Sigma-phi-chi}, the WT
operator \bref{Xi} has contributions from the IR regulator.

Let us integrate over the PV fields in \bref{Gamma-Gamma}.  Neglecting
terms of the order $\left<O(\chi^4)\right>_{\chi}\sim O(\hbar^2)$, we find that
\bea
\left<\Xi_{\rm tot}\right>_{\chi} &=& \Xi_{\rm reg}[\phi,~\phi^{*}]
=\sum_{i=1}^{3} \Xi^{\{i\}}_{\rm reg}[\phi,~\phi^{*}],\nn\\
 \Xi^{\{1\}}_{\rm reg}[\phi,\phi^{*}]&\equiv&
\frac{1}{2}\left(S_{k},~ S_{k}\right)_{\phi},\nn\\
\Xi^{\{2\}}_{\rm reg}[\phi,\phi^{*}]&\equiv&\frac{1}{2} \left<
\left(\phi^{A} f_{k}^{2} R^{k}_{AB}\phi^{B},~S_{\rm
PV}\right)_{\phi}\right>_{\chi}\nn\\
&=& -\frac{\hbar}{2} \phi^{C} f_{k}^{2}
(R^{k})_{CA}(-)^{\ep_{B}}\left(\frac{\partial^l}{\partial \phi^{*}_{A}}
[D  + \Lambda T]_{BE}\right)
\left([D  + \Lambda T]^{-1}\right)^{BE}\nn\\
&=& -\frac{\hbar}{4} ~
\left(\phi^{A}f_{k}^{2}R^{k}_{AB}\phi^{B},~{\rm str}\ln \left[T{\cal O}+
T\Lambda\right]\right)_{\phi}, \nn\\
\Xi^{\{3\}}_{\rm reg}[\phi,\phi^{*}]&\equiv& \left< \left(S^{\Lambda}_{\rm PV},~S^{(0)}\right)_{\phi} + \left(S^{\Lambda}_{\rm PV},~S^{(0)}_{\rm PV}\right)_{\chi}\right>_{\chi}\nn\\
&=&  \Lambda \left<\chi^{A}(T{\cal K})_{AB}\chi^{B}\right>_{\chi}= 
-\hbar~{\rm tr}\left({\cal K}[{1+ {\cal O}/\Lambda}]^{-1}\right).
\label{chi-integ2}
\eea

In contrast to the previous subsection, we need to perform the remaining
$\phi$ integration to separate the IR regulator:
\bea
\frac{1}{2}\left({\hat \Gamma}_{k\Lambda}, {\hat \Gamma}_{k\Lambda}\right)_{\vf}
&=&  \frac{\int {\cal D}\phi ~ \Xi_{\rm reg}[\phi,~\phi^{*}]
\exp\Bigl\{ -\left
(S_{k}[\phi,~\phi^{*}]-
\frac{\partial^{r}{\hat \Gamma}_{k\Lambda}}{\partial \vf^{A}}
(\phi-\vf)^{A}\right)/\hbar \Bigr\}}  
{\int {\cal D}\phi \exp\Bigl\{ -\left
(S_{k}[\phi,~\phi^{*}]-
\frac{\partial^{r}{\hat \Gamma}_{k\Lambda}}{\partial \vf^{A}}
(\phi-\vf)^{A}\right)/\hbar \Bigr\}}\nn\\ 
&\equiv&  \left<\Xi_{\rm reg}[\phi,\phi^{*}]\right>_{\phi}. 
\label{phi-int}
\eea
Making a shift of variables $\phi^{A} \to \phi^{A}+ \vf^{A}$, and
an expansion in powers of $\phi^{A}$, one obtains
\bea
\frac{1}{2}\left({\hat \Gamma}_{k\Lambda}, {\hat \Gamma}_{k\Lambda}\right)_{\vf}
&=& \left<\Xi_{\rm reg}[\vf,\vf^{*}] +  \phi^{A}
\frac{\partial^l~ \Xi_{\rm reg}}{\partial \vf^{A}}+ \frac{1}{2} \phi^{A}
\frac{\partial^l}{\partial \vf^{A}}\frac{\partial^r ~\Xi_{\rm reg}}
{\partial \vf^{B}} \phi^{B}+ \cdots\right>_{\phi}\nn\\
&=& \sum_{i=1}^{4} \Xi^{\{i\}}_{\rm reg}[\vf,~\vf^{*}]
\label{phi-int2}
\eea
where $ \Xi^{\{i\}}_{\rm reg}[\vf,~\vf^{*}]~~(i=1,~3)$ are the same as
those in \bref{chi-integ2} with fields and antifields replaced by
$\{\vf,~\vf^{*}\}$. There appears a new term from the $\phi$
integration. To lowest order of $\hbar$, it is give by
\bea
 \Xi^{\{4\}}_{\rm reg}[\vf,~\vf^{*}]&=&\frac{\hbar}{2}
\frac{\partial^l}{\partial \vf^{A}}\frac{\partial^r \Xi_{\rm reg}^{\{1\}}}
{\partial \vf^{B}}
\left(\left[f_{k}^{2}R^{k}+ T{\cal O}\right]^{-1}\right)^{BA}\nn\\
&=&\frac{\hbar}{2}\frac{\partial^l}{\partial \vf^{A}}
\frac{\partial^r}{\partial \vf^{B}}
\left(\vf^{C}f_{k}^{2}R^{k}_{CE}\vf^{E},~S^{(0)}\right)_{\vf}\left(\left[f_{k}^{2}R^{k}+
T{\cal O}\right]^{-1}\right)^{BA}\nn\\
 &=& \hbar \left( f_{k}^{2}R^{k}_{AC}K^{C}_{~B} 
+\frac{1}{2} (-)^{\ep_A}\vf^{C}f_{k}^{2}R^{k}_{CE}\frac{\partial^l}{\partial
\vf_{E}^{*}}D_{AB}\right)\left(\left[f_{k}^{2}R^{k}+ T{\cal
O}\right]^{-1}\right)^{BA}\nn\\
 &=&  \hbar~ f_{k}^{2}R^{k}_{AC}K^{C}_{~B}\left( \left[f_{k}^{2}R^{k}
+ T{\cal O}\right]^{-1}\right)^{BA}\nn\\
&{}& + 
 \frac{\hbar}{4} ~
\left(\vf^{A}f_{k}^{2}R^{k}_{AB}\vf^{B},~{\rm str}\ln \left[f_{k}^{2}R^{k}
+ T{\cal O}\right]\right)_{\vf}.
\label{Sigma4}
\eea
Combining this with \bref{chi-integ2} gives
\bea
&{}&\hspace{-7mm}\frac{1}{2}\left({\hat \Gamma}_{k\Lambda}, {\hat \Gamma}_{k\Lambda}\right)_{\vf}
= \frac{1}{2}\left(S_{k},~S_{k}\right)_{\vf}+\hbar f_{k}^{2}R^{k}_{AC}K^{C}_{~B}\left( \left[f_{k}^{2}R^{k}
+ T{\cal O}\right]^{-1}\right)^{BA}
\hspace{-4mm}
-\hbar~{\rm tr}\left({\cal K}[{1+ {\cal O}/\Lambda}]^{-1}\right)\nn\\
&{}& ~~~~~~~~~~~~~+ \frac{\hbar}{4} ~
\left(\vf^{A}f_{k}^{2}R^{k}_{AB}\vf^{B},~{\rm str}\ln \left[f_{k}^{2}R^{k}
+ T{\cal O}\right]- {\rm str}\ln \left[T{\cal O}+
T\Lambda\right]\right)_{\vf}.
\label{G-G5}
\eea
Using 
\bea
&{}&\Gamma_{k\Lambda}={\hat \Gamma}_{k\Lambda}-\frac{1}{2} 
\vf^{A}f_{k}^{2}R^{k}_{AB}\vf^{B}=\Gamma^{(0)} +
\hbar\Gamma_{k\Lambda}^{(1)},\nn\\
&{}&\Gamma^{(0)}= S^{(0)}, ~~~~~~~~\Gamma_{k\Lambda}^{(1)}= S^{(1)}+
{\rm str}\ln \left([{f_{k}^{2}R^{k}+ T{\cal O}}][{T{\cal O}+ T\Lambda}]^{-1}\right),\nn\\   
&{}&\left(\Delta_{\vf}S^{(0)}\right)_{\rm reg}= 
{\rm tr} \left({\cal K}[{1+ {\cal O}/\Lambda}]^{-1}\right),
\label{WTfinal}
\eea
one finally obtains
\bea
&{}&\frac{1}{2}\left({\Gamma}_{k\Lambda}, {\Gamma}_{k\Lambda}\right)_{\vf}- \hbar f_{k}^{2}R^{k}_{AC} \left(\frac{\pa^{l} \pa^{r} {\G}^{(0)}}
{\pa \vf_{C}^{*} \pa \vf^{B}}\right)
\Biggl(\left[f_{k}^{2} R^{k}+
\frac{\pa^{l} \pa^{r} {\G}^{(0)}}{\pa \vf \pa\vf}\right]^{-1}\Biggr)^{BA}\nn\\ 
&{}&~~~~~~~~~~~~~~~= \Sigma_{\rm reg}[\vf,~\vf^{*}]\equiv \hbar \left(S^{(1)},~S^{(0)}\right)_{\vf}-
\hbar~\left(\Delta_{\vf}S^{(0)}\right)_{\rm reg}.
\label{WTfinal2}
\eea
The rhs is again the regularized WT operator for the UV action.  

The regularized WT operator in \bref{one-loop-result} and
\bref{WTfinal2} contains $(\Delta S^{(0)})_{\rm reg}$.  This is exactly
the TVV jacobian factor associated with the BRS transformation in the
original gauge-fixed UV action. Thus, our main task for solving the QME
is to compute the TVV jacobian factor for a given classical UV
action. For an anomaly free theory, the TVV jacobian factor becomes a
coboundary term, and one can find a local counter action $S^{(1)}$ for
which $\Sigma_{\rm reg}|_{\Lambda \to \infty}=0$. The QME for the UV
action is solved in this way.  These calculations will be done
explicitly for the pure Yang-Mills theory to exemplify the procedure.

When there exists a non-trivial anomaly, such a local counter action
cannot be constructed and $\Sigma_{\rm reg}|_{\Lambda \to \infty}( \neq
0)$ corresponds to the BRS (gauge) anomaly. It follows from $\Sigma_{k}
=\Sigma_{\rm reg}|_{\Lambda \to \infty} \neq 0 $ that, when expressed by
the classical field configuration, the BRS anomaly does not depend on
the IR cutoff and persists along the RG flow.

%%%%%%%%%%%%%%%%%%%%%%%%%%%%%%%%%%%%%%%%%%%%%%%%%%%%%%%%%%%%%%%%%%%%%
%%%%%%%%%%%%%%%%%%%%%%%%%%%%%%%%%%%%%%%%%%%%%%%%%%%%%%%%%%%%%%%%%%%%%
%%%%%%%%%%%%%%%%%%%%%%%%%%%%%%%%%%%%%%%%%%%%%%%%%%%%%%%%%%%%%%%%%%%%%
\section{The quantum master equation in pure Yang-Mills theory}

Here we construct the quantum UV action for the four-dimensional SU(N)
pure Yang-Mills theory based on the TVV formalism given in
ref. \cite{TVV}.  The TVV jacobian factor and the local counter terms
for cancelling superficial anomaly terms was already calculated in
ref. \cite{DeJonghe}.  Since the calculation is somewhat tedious, we
give here some results including those obtained at intermediate steps
which were not given in ref. \cite{DeJonghe}. We also retain the
contributions from the antifields.

We begin with a gauge-fixed UV action in the Feynman gauge
\bea
S^{(0)}[\vf,\vf^{*}] &=& - {\rm tr} \int d^{4}x
\Biggl[\frac{1}{4}F_{\mu\nu}^{2} + \frac{1}{2} (\partial \cdot A)^{2}-
{\bar C}\partial \cdot D~C 
\nn\\
&{}&~~~~~+ A^{*}\cdot D ~C + C^{*} ~C^{2} - {\bar C}^{*}~\partial \cdot 
A
\Biggr],
\label{pure-YM-action}
\eea
where $B$ fields are eliminated and the trace $\rm tr$ is over gauge
indices.  The covariant derivative is given as $D_{\rho} C =
\pa_{\rho} C + [A_{\rho},~C]$.  For the fields $\vf^{A}=\{{\bar
C}(x),~A_{\mu}(x),~C(x)\}$ and their antifields, the basic matrices we
need are given by
\bea
L_{AB}&=&\frac{\partial^l}{\partial \vf^{A}}\frac{\partial^r S}{\partial \vf^{B}}=
\left(
\begin{array}{ccc}
0 & \partial_{\nu}C & - \partial\cdot D\\ 
- C\partial_{\mu} & R_{\mu\nu} & -( \partial_{\mu}{\bar C})- A^{*}_{\mu}\\
D\cdot \partial & -( \partial_{\nu}{\bar C})- A^{*}_{\nu}&0
\end{array}
\right)(x)~\delta(x-y),\nn\\
R_{\mu\nu} &=& D_{\mu}D_{\nu}- \delta_{\mu\nu}D^2 -
 \pa_{\mu}\pa_{\nu}-2 F_{\mu\nu},\nn\\
T_{AB} &=& 
\left(
\begin{array}{ccc}
0 & 0 & 1\\ 
0 &  \delta_{\mu\nu} & 0\\
-1 & 0 & 0
\end{array}
\right)(x)~\delta(x-y),\label{marix1}\\
{\cal K}^{B}_{~A} &=& K^{B}_{~A}= \frac{\partial^l}{\partial \vf^{*}_{B}}\frac{\partial^r S}{\partial \vf^{A}}=
\left(
\begin{array}{ccc}
0 & -\partial_{\mu} & 0\\ 
0 & - \delta^{\nu}_{\mu}C & D^{\nu}\\
0 & 0 & C
\end{array}
\right)(y)~\delta(y-x).\nn\\
\eea
In order to use the heat kernel method, we take the exponential form of the trace,
\bea
&{}&\left(\Delta_{\vf}S^{(0)}\right)_{\rm reg} =
 \left(K_{s}\right)^{B}_{~A}\left(\exp \left(- {\cal O}/M^{2}\right)\right)^{A}_{~B}\nn\\
&{}&~~~=\int d^{4}x \int \frac{d^{4}k}{(2\pi)^{4}}{\rm tr}\left[K_{s}(x)
\exp (-ik\cdot x)~\exp\left(- {\cal O}(x)/M^{2}\right)\right]
\exp (ik\cdot x),
\label{exp1}
\eea
where $\Lambda=M^2$ and 
\bea
\left(K_{s}\right)^{B}_{~A} &=& \frac{1}{2}(K - T^{-1}K^{t}T)^{B}_{~A}=\frac{1}{2}
\left(
\begin{array}{ccc}
-C & A_{\mu} & 0\\ 
0 & 0 & A^{\nu}\\
0 & 0 & C
\end{array}
\right)(y)~\delta(y-x)= K_{s}(y)~\delta(y-x),\nn\\
{\cal O}^{A}_{~B} &=& \left(T^{-1}L\right)^{A}_{~B}=
\left(
\begin{array}{ccc}
- D \cdot \partial & (\partial_{\nu}{\bar C}) + A^{*}_{\nu}& 0\\ 
-C \partial^{\mu} & R^{\mu}_{~\nu} & -(\partial^{\mu}{\bar C})- A^{*\mu} \\
0 &\partial_{\nu}C  & -\partial \cdot D 
\end{array}
\right)(x)~\delta(x-y)\nn\\
&{}&~~~~~~~~~~~~~~~~~ ={\cal O}(x)~\delta(x-y),\nn\\
\label{matrix2}
\eea
The use of the ``symmetrized'' matrix \cite{DeJonghe} $K_{s}$ defined
with the transposed matrix $K^{t}$ simplifies our trace
calculation. We decompose the matrix ${\cal O}(x)$ as
\bea
{\cal O}(x) = -\left(\partial_{\rho}{\bf 1}+ Y_{\rho} \right)
\delta^{\rho\sigma}\left(\partial_{\sigma}{\bf 1}+ Y_{\sigma}\right) -E,
\label{decom}
\eea
where ${\bf 1},~Y_{\rho},~E$ are $6 \times 6$ matrices.  These matrices are given by
\bea
Y_{\rho} &=& A_{\rho}{\bf 1}+ \frac{1}{2}
\left(
\begin{array}{ccc}
-A_{\rho} & 0 & 0\\ 
C \delta^{\mu}_{\rho} & - A^{\mu}\delta_{\nu\rho}- A_{\nu}\delta^{\mu}_{\rho} & 0\\
0 & -C\delta_{\rho\nu}  & -A_{\rho}
\end{array}
\right),\nn\\
E &=& \frac{1}{4}A^{2}{\bf 1} + 
\left(
\begin{array}{ccc}
-\frac{1}{2}\partial\cdot A & - \partial_{\nu}{\bar C} - A^{*}_{\nu}& 0\\ 
V^{\mu} & E^{\mu}_{~\nu} &\partial^{\mu}{\bar C}+ A^{*\mu} \\
C^{2} & - V^{t}_{\nu}  & \frac{1}{2}\partial\cdot A
\end{array}
\right),
\label{matrix3}\\
V^{\mu} &=& -\frac{1}{2}\partial^{\mu} C +
\frac{1}{4}\left(3A^{\mu}C-CA^{\mu}\right), ~~~~~ - V^{t}_{\nu}=  -\frac{1}{2}
\partial_{\nu} C  +
\frac{1}{4}\left(A_{\nu}C-3~CA_{\nu}\right),\nn\\
E_{\mu\nu} &=& \frac{3}{2}\left(\partial_{\mu}A_{\nu}- 
\partial_{\nu}A_{\mu}\right)+\frac{1}{2}A_{\mu}A_{\nu}-A_{\nu}A_{\mu}.\nn    
\eea
As shown in Appendix B, large $M$ expansion of the momentum integration
in \bref{exp1} becomes 
\bea
&{}&\int \frac{d^{4}k}{(2\pi)^{4}}
\exp (-ik\cdot x)~\exp\left(- {\cal O}(x)/M^{2}\right)
\exp (ik\cdot x)\nn\\
&{}&=-\frac{M^2}{(4\pi)^2}{\rm tr}(A \cdot \pa C) +   
\frac{1}{6(4\pi)^2}{\rm tr}\biggl[C
\pa^2 \pa\cdot A - 2 (\pa C)\cdot A \pa\cdot A + 2 (\pa^{\rho}C)A^{\s}\pa_{\rho}A_{\s} \nn\\
&{}&\quad ~~~~~~~~~~~ -2(\pa^{\rho}C)A^{\s}\pa_{\s}A_{\rho}
+2(\pa^{\rho}C)A^{\s}A_{\rho}A_{\s} \biggr] + O(1/M^2).
\label{mom-int1} 
\eea 
We notice that the rhs of \bref{mom-int1} contains neither the
antighosts nor the antifields, and is shown to be a coboundary
term. Thus, we find that the QME is satisfied in the $M \to \infty$ limit 
\bea
\left(\Delta_{\vf}S^{(0)}\right)_{\rm reg}= - \left(S^{(0)},~S^{(1)}\right)_{\vf},
\eea
with the counter action\footnote{The conventional gauge invariant
counter terms should be added to construct the total counter action.
These terms depend on the renormalization conditions.  We 
do not further discuss this point in this paper.}
\bea
S^{(1)} &=& \int d^4 x \frac{1}{24(4\pi)^2}{\rm tr}\biggl[24 M^2 A^2 +
(\pa_{\mu}A_{\rho})^2 - 3 (\pa\cdot A)^2 -4
A_{\mu}(\pa_{\rho}A^{\mu})A^{\rho}\nn\\
 &{}& + (A^2)^2 - 3 A_{\mu} A_{\nu}A^{\mu}A^{\nu}\biggr].
\label{counter-action}
\eea
The finite part of \bref{counter-action} is the same as that given in
ref. \cite{DeJonghe}.  

With the counter action \bref{counter-action}, $S^{(0)} + {\hbar}
S^{(1)}$ is a UV action satisfying the QME at the one-loop level.
It is quite interesting to realize that the IR theory generated from the 
UV action has the exact renormalized BRS symmetry, though the counter
action itself breaks the gauge symmetry.  

%%%%%%%%%%%%%%%%%%%%%%%%%%%%%%%%%%%%%%%%%%%%%%%%%%%%%%%%%%%%%%%%%%%%%
%%%%%%%%%%%%%%%%%%%%%%%%%%%%%%%%%%%%%%%%%%%%%%%%%%%%%%%%%%%%%%%%%%%%%
\section{Discussion}
%%%%%%%%%%%%%%%%%%%%%%%%%%%%%%%%%%%%%%%%%%%%%%%%%%%%%%%%%%%%%%%%%%%%%
%%%%%%%%%%%%%%%%%%%%%%%%%%%%%%%%%%%%%%%%%%%%%%%%%%%%%%%%%%%%%%%%%%%%%

In our results \bref{one-loop-result} and \bref{WTfinal2}, the rhs of
these relations can be computed using the UV action, independently of
the IR regularization. Therefore, whether the renormalized symmetry
along the RG flow exists or not can be determined solely from the UV
action. This consequence should be compared with the previous analysis
based on the fine-tuning \cite{Bonini0,Bonini1}.  The fine-tuning
analysis corresponds to the computation of the lhs of \bref{WTfinal2}
using a single regulator for both IR and UV regularizations for fixed
$k$.  We would like to emphasize the following advantage of studying the
rhs rather than the lhs: The presence of the effective BRS symmetry
along the RG flow is guaranteed by properties of the UV action without
reference to the IR regularization.

Let us make a remark on application of our formalism to global
symmetries such as chiral symmetry.  For a given global symmetry, the
equations given in the previous section are valid except that ghosts
associated with the symmetry are constant.  In order to discuss a
possible anomaly, we may consider a given theory in a compactified space
where the boundary effects can be taken into account. Alternatively, we
may introduce space-time dependent ghost fields instead of constant
fields. Then, the anomaly is identified with $\Sigma_{\rm reg}|_{\Lambda
\to \infty}$ or its functional derivative with respect to the ghosts.

In the forthcoming paper \cite{Igarashi2}, we apply our formalism to
global symmetries, such as the chiral and SU(N) flavor symmetries,
toward our goal to provide a formulation for a non-perturbative study.

%%%%%%%%%%%%%%%%%%%%%%%%%%%%%%%%%%%%%%%%%%%%%%%%%%%%%%%%%%%%%%%%%%%%%
%%%%%%%%%%%%%%%%%%%%%%%%%%%%%%%%%%%%%%%%%%%%%%%%%%%%%%%%%%%%%%%%%%%%%
\section*{Acknowledgments} 

This work is supported in part by the Grants-in-Aid for Scientific
Research No. 12640258, 12640259, and 13135209 from the Japan Society for
the Promotion of Science.

%%%%%%%%%%%%%%%%%%%%%%%%%%%%%%%%%%%%%%%%%%%%%%%%%%%%%%%%%%%%%%%%%%%%%
%%%%%%%%%%%%%%%%%%%%%%%%%%%%%%%%%%%%%%%%%%%%%%%%%%%%%%%%%%%%%%%%%%%%%
%%%%%%%%%%%%%%%%%%%%%%%%%%%%%%%%%%%%%%%%%%%%%%%%%%%%%%%%%%%%%%%%%%%%%
\appendix
%%%%%%%%%%%%%%%%%%%%%%%%%%%%%%%%%%%%%%%%%%%%%%%%%%%%%%%%%%%%%%%%%%%%%

\section{Derivation of regularized WT identity \bref{Gamma-Gamma}}
%%%%%%%%%%%%%%%%%%%%%%%%%%%%%%%%%%%%%%%%%%%%%%%%%%%%%%%%%%%%%%%%%%%%%
The WT identity \bref{Gamma-Gamma} can be derived as follows.  We
consider the partition function \bref{reg-what} with integration
variables replaced by
\bea
\phi^{\prime A} & =& \phi^{A} + (\phi^{A}, S_{\rm tot})_{\phi} \lambda, \nn\\
\chi^{\prime A} & =& \chi^{A} + (\chi^{A}, S_{\rm tot})_{\chi} \lambda,
\label{change-var}
\eea
where  $(~,~)_{\chi}$ is the antibracket with respect to the PV fields, and
$\lambda$ is an anticommuting constant. The infinitesimal change of
variables leads to 
\bea
\int {\cal D} \phi {\cal D} \chi {\cal D} \chi^{*}\prod_{B}\delta
(\chi^{*}_{B})\exp\Bigl\{-
\left(S_{\rm tot}-j_{C}\phi^{C}\right)/\hbar \Bigl\}
~~~~~~~~~~~~~~~~~~~~~~~~~~~~~~~~~~~~~~ \nn\\
\times \left\{\frac{\partial^{r} S_{\rm tot}}{\partial \phi^{A}}
(\phi^{A}, S_{\rm tot})_{\phi} + \frac{\partial^{r} S_{\rm tot}}{\partial \chi^{A}}
(\chi^{A}, S_{\rm tot})_{\chi} - \hbar \left(\Delta_{\phi}+\Delta_{\chi}\right)
S_{\rm tot}- j_{A}(\phi^{A}, S_{\rm tot})_{\phi}\right\}=0.
\label{WT1}
\eea
The second line apart from the last term is $2~\Xi_{\rm tot}$. Let us
rewrite the contribution from the last term in \bref{WT1} by using the
derivatives of the Legendre effective action \bref{Ghat2}:
\bea
\frac{\partial^{r}{\hat \Gamma}_{k}}{\partial \vf^{A}} &=& 
j_{A}[\vf,\vf^{*}],\nn\\
\frac{\partial^{l}\hat{\Gamma}_{k}}{\partial \vf_{A}^{*}} &=& 
- \hbar {\hat Z}^{-1} \frac{\partial^{l}{\hat Z}}{\partial \vf_{A}^{*}}
+ \frac{\partial^{l}j_{B}}{\partial \vf_{A}^{*}}\vf^{B},
\label{Gamma-der} 
\eea
where 
\bea
\hbar  \frac{\partial^{l}{\hat Z}}{\partial \vf_{A}^{*}}&=& N_{k}\int {\cal D} \phi {\cal D} \chi {\cal D} \chi^{*}\prod_{A}\delta
(\chi^{*}_{A})\exp \left(j_{B}\phi^{B}/\hbar\right)
\nn\\ &{}& ~~~~~~~ \times 
\left\{\frac{\partial^{l}j_{B}}{\partial \vf_{A}^{*}}\phi^{B}+
\hbar\frac{\partial^{l}}{\partial \vf_{A}^{*}} \right\} \exp \left(-S_{\rm tot}/\hbar\right).  
\label{Z-vf}
\eea
Then, it follows that
\bea
j_{A}\frac{\partial^l{\hat\Gamma}_{k}}{\partial \vf_{A}^*} &=& 
- \hbar {\hat Z}^{-1}\int {\cal D} \phi {\cal D} \chi {\cal D} 
\chi^{*}\prod_{B}\delta(\chi^{*}_{B}) \exp
\left(j_{B}\phi^{B}/\hbar\right)
j_{A}\frac{\partial^l}{\partial\phi_{A}^*}
\exp(-S_{\rm tot}/\hbar)
\nn\\
&=& \int {\cal D} \phi {\cal D} \chi {\cal D} 
\chi^{*}\prod_{B}\delta(\chi^{*}_{B})j_{A}(\phi^{A}, S_{\rm tot})_{\phi}
\exp\Bigl\{-\left(S_{\rm tot}-j_{C}\phi^{C}\right)/\hbar \Bigr\}.  
\label{j-Gamma}
\eea
Using \bref{WT1} and \bref{j-Gamma}, one obtains
\bea
\frac{1}{2}\left({\hat\Gamma}_{k},~{\hat\Gamma}_{k}\right)_{\varphi}= {\hat Z}^{-1}
\int {\cal D} \phi {\cal D} \chi {\cal D} 
\chi^{*}\prod_{B}\delta(\chi^{*}_{B})~\Xi_{\rm tot}\exp\Bigl\{-\left(S_{\rm
tot}-j_{A}\phi^{A}\right)/\hbar \Bigr\},
\eea
which gives \bref{Gamma-Gamma}.

\section{Computation of the TVV jacobian factor \bref{mom-int1}}
We first consider the momentum integration in \bref{exp1}, using the
decomposition \bref{decom} and the notation ${\cal D}_{\rho}= \pa_{\rho} 
+ Y_{\rho}$:
\bea
&{}&\int \frac{d^{4}k}{(2\pi)^{4}}
\exp (-ik\cdot x)~\exp\left(- {\cal O}(x)/M^{2}\right)
\exp (ik\cdot x) \nn\\
&{}& ~~= \int \frac{d^{4}k}{(2\pi)^{4}} 
\left(\exp\left[ \frac{1}{M^2}(\partial_{\rho} + i k_{\rho} + Y_{\rho} +E)\right]\right) \cdot {\bf 1}\nn\\
&{}&~~= \int \frac{d^{4}k}{(2\pi)^{4}} M^4 \exp\left[-k^2 + \frac{2ik^{\rho}}{M}{\cal D}_{\rho} + \frac{1}{M^2}({\cal D}^2 + E)\right] \cdot {\bf 1}\nn\\
&{}&~~= M^{2}~a(x) +~b(x)+~O(1/M^2)
\label{k-int1}
\eea
where the replacement $k \to k~M$ is made to get the third line, and the large 
$M$ limit is taken in the last line.  
Using the Euclidean integral
\bea
\int d^{4}k (k^{2})^{n} \exp{(-g~ k^{2})} 
= \pi^{2} (n+1)! ~g^{-(n+2)}~~~~~~~~~~(g >0),
\label{mom-int}
\eea
we can show that the matrix $a(x)$ takes of the form
\bea
a(x) &=& \int \frac{d^{4}k}{(2\pi)^{4}} \exp(-k^2)
\biggl[\frac{(2i)^{2}}{2!}k^{\rho}k^{\sigma}{\cal D}_{\rho}{\cal D}_{\sigma}+ 
({\cal D}^2 + E)\biggr]=\frac{1}{(4\pi)^2}(2{\cal D}^2 + E)\nn\\
&=& \frac{1}{(4\pi)^2}(2 \pa \cdot Y + 2 Y^{2} + E).
\label{a-rep}
\eea
This leads to 
\bea
M^{2}{\rm tr}\left(K_{s}(x)~a(x)\right)=- \frac{M^2}{(4\pi)^2}{\rm tr}
(A \cdot \pa C),
\label{M^2-term2}
\eea
where the matrix $K_{s}$ given in \bref{matrix2} is used, and total derivative terms are ignored.  For $b(x)$, we have
\bea
b(x) &=& 
\int \frac{d^{4}k}{(2\pi)^{4}} \exp(-k^2)~\biggl[\frac{(2i)^{4}}{4!} 
k^{\rho}k^{\sigma}k^{\mu}k^{\nu}{\cal D}_{\rho}{\cal D}_{\sigma}{\cal D}_{\mu}
{\cal D}_{\nu} \nn\\ 
&{}&~~+\frac{(2i)^{2}}{3!} k^{\rho}k^{\sigma}\{{\cal D}_{\rho}{\cal D}_{\sigma}({\cal D}^2 + E) + {\cal D}_{\rho}({\cal D}^2 + E){\cal D}_{\sigma} +({\cal D}^2 + E) {\cal D}_{\rho}{\cal D}_{\sigma}\}+ \frac{1}{2!}({\cal D}^2 + E)\biggr]
\nn\\
 &=& \frac{1}{(4\pi)^2} 
 \left[\frac{1}{6}{\cal D}^{\rho}{\cal D}^{\sigma}
 ({\cal D}_{\rho}{\cal D}_{\sigma}-{\cal D}_{\sigma}{\cal D}_{\rho}) + \frac{1}{2}E^2 + \frac{1}{6}({\cal D}^2 E +E {\cal D}^2 -2 {\cal D}^{\rho}
 E{\cal D}_{\rho})\right]\nn\\
&=& \frac{1}{(4\pi)^2} 
\left(\frac{1}{12} W_{\rho\sigma}^2 + \frac{1}{2} E^2 + \frac{1}{6} 
\nabla_{\rho} \nabla^{\rho} E \right)\nn\\
&=& \frac{1}{24(4\pi)^2}
\left(
\begin{array}{ccc}
\alpha  & * & *\\ 
\beta^{\mu} & * & * \\
{} * & \gamma_{\nu}  & \eta 
\end{array}
\right)(x),
\label{b-rep}
\eea
where $\nabla^{\rho}E \equiv \partial^{\rho} E + [Y^{\rho},E]$.
$\alpha, ~\beta^{\mu},~\gamma_{\nu}$, and $ \eta$ are the matrix
elements needed for our trace calculation.  Other matrix elements
denoted by the asterisks do not contribute to the trace.  Actually, we find that
\bea
{\rm tr}\left(K_{s}(x)~b(x)\right) &=& \frac{1}{24(4\pi)^2}{\rm tr}\left[C(\eta - \alpha) + (A_{\mu}\beta^{\mu} + A^{\nu}\gamma_{\nu})\right]\nn\\
&=& \frac{1}{12(4\pi)^2}{\rm tr}\left(-C\alpha + A_{\mu}\beta^{\mu}\right),
\label{M^0-term2}
\eea
because of symmetry property
\bea
{\rm tr}(C\eta) = -{\rm tr}(C \alpha),~~~~~~~~
{\rm tr} (A_{\mu}\beta^{\mu}) = {\rm tr}(A^{\nu}\gamma_{\nu}).
\label{symm-rel}
\eea

The matrix elements $\alpha$ and $\beta^{\mu}$ are given by
\bea
\alpha &=& \alpha(W^2) + 6 \alpha(E^2) + 2 \alpha
(\nabla_{\rho} \nabla^{\rho} E),\nn\\
\beta^{\mu} &=& \beta^{\mu}(W^2) + 6 \beta^{\mu}(E^2) + 
2 \beta^{\mu}(\nabla_{\rho} \nabla^{\rho} E),
\label{matrix-ele}
\eea
where $\alpha(W^2)$, for example, denotes 
the contribution from the matrix $W^2$.  We decompose further the last term 
in \bref{b-rep} into 
\bea
\nabla_{\rho} \nabla^{\rho} E &=& \pa^{\rho}(\pa_{\rho}E + Y_{\rho}E - 
EY_{\rho}) +[Y^{\rho}, \nabla_{\rho}E]\nn\\
&=& \pa^2 E + (\pa\cdot Y)E +2Y\cdot \pa E - 2(\pa E)\cdot Y \nn\\
&{}&- E\cdot\pa Y + Y^2 E -2 Y^{\rho}E Y_{\rho} + E Y^2.
\label{nabla-E}
\eea
{}For the matrices given above, $\alpha$'s are given by
\bea
\alpha(W^2) &=& \frac{1}{4}\left(F_{\rho\sigma}- \frac{1}{2}[A_{\rho},A_{\sigma}]\right)^{2},\nn\\
\alpha(E^2) &=& \left(\frac{1}{4}A^2 +\frac{1}{2}\pa\cdot A\right)^2 - 
\left(A^{*}_{\rho} + \pa_{\rho}{\bar C} \right)V^{\rho},\nn\\
\alpha(\pa^2 E) &=& - \left(\frac{1}{4}\pa^2 A^2 +\frac{1}{2}\pa^2 \pa\cdot A\right)
,\nn\\
\alpha((\pa\cdot Y)E) &=& - \frac{1}{2}(\pa\cdot A)\left(\frac{1}{4}A^2 +\frac{1}{2}\pa\cdot A\right),
\label{alpha}\\
\alpha(2Y\cdot \pa E) &=& - A\cdot\pa \left(\frac{1}{4}A^2 +\frac{1}{2}\pa\cdot A\right),\nn\\
\alpha(-2(\pa E)\cdot Y) &=& \left[\pa^{\rho}\left(\frac{1}{4}A^2 +\frac{1}{2}\pa\cdot A\right)\right]A_{\rho} + \pa\cdot(A^* + \pa {\bar C} ) C ,\nn\\
\alpha(- E\cdot\pa Y) &=& \frac{1}{2}\left(\frac{1}{4}A^2 +\frac{1}{2}\pa\cdot A\right)\pa\cdot A + \frac{1}{2}(A^* + \pa {\bar C})\cdot \pa C ,\nn\\
\alpha(Y^2 E) &=& -\frac{A^2}{4}\left(\frac{1}{4}A^2 +\frac{1}{2}\pa\cdot A\right) ,\nn\\
\alpha(-2 Y^{\rho}E Y_{\rho}) &=& \frac{1}{2}A^{\rho}
\left[\left(\frac{1}{4}A^2 +\frac{1}{2}\pa\cdot A\right)A_{\rho}+ (A^{*}_{\rho}
+ \pa_{\rho}{\bar C})C\right],\nn\\
\alpha(EY^2) &=& - \frac{1}{4}\left(\frac{1}{4}A^2 +\frac{1}{2}\pa\cdot A\right) A^2 + \frac{1}{4}(A^{*}_{\rho} + \pa_{\rho}{\bar C})(3 A^{\rho}C-C A^{\rho}).
\nn 
\eea
Likewise, we have
\bea
\beta^{\mu}(W^2) &=& \frac{1}{2}\left(D_{\rho}C \delta^{\mu}_{~\s}+\frac{1}{2}
(C A_{\rho}+ A_{\rho}C)\delta^{\mu}_{~\s}\right) 
\left(F_{\rho\sigma}- \frac{1}{2}[A_{\rho},A_{\sigma}]\right) \nn\\
&{}& + R^{\mu\nu,\rho\s}
\left(D_{\rho}C \delta_{\nu\s}+\frac{1}{2}
(C A_{\rho}+ A_{\rho}C)\delta_{\nu\s}\right),\nn\\
\beta^{\mu}(E^2) &=&- V^{\mu}
\left(\frac{1}{4}A^2 +\frac{1}{2}\pa\cdot A\right)
+ \left(E^{\mu}_{~\rho}- \frac{A^2}{4}\delta^{\mu}_{~\rho}\right)V^{\rho}
+ \left(A^{*\mu} + \pa^{\mu}{\bar C}\right)C^2 ,\nn\\
\beta^{\mu}(\pa^2 E) &=& \pa^2 V^{\mu}
,\nn\\
\beta^{\mu}((\pa\cdot Y)E) &=& - \frac{1}{2}(\pa^{\mu} C) 
\left(\frac{1}{4}A^2 +\frac{1}{2}\pa\cdot A\right)
+ \frac{1}{2}(2 \pa\cdot A \delta^{\mu}_{~\rho} - \pa_{\rho} A^{\mu} - 
\pa^{\mu} A_{\rho})V^{\rho}
,\nn\\
\beta^{\mu}(2Y\cdot \pa E) &=& - C \pa^{\mu}\left(\frac{1}{4}A^2 +\frac{1}{2}\pa\cdot A\right) + (2 A \cdot \pa \delta^{\mu}_{~\rho} - A^{\mu}\pa_{\rho}- A_{\rho}
\pa^{\mu})V^{\rho},\nn\\
\beta^{\mu}(-2(\pa E)\cdot Y) &=& - (\pa_{\rho}V^{\mu})A^{\rho} - 
\left[\pa^{\rho}\left(E^{\mu}_{~\rho} - \frac{A^2}{4}\delta^{\mu}_{~\rho}\right)
\right] C,
\label{beta-mu}\\
\beta^{\mu}(- E\cdot\pa Y) &=&- \frac{1}{2} V^{\mu} \pa\cdot A -\frac{1}{2}
\left(E^{\mu}_{~\nu} - \frac{A^2}{4}\delta^{\mu}_{~\nu}\right) \pa^{\nu} C,\nn\\
\beta^{\mu}(Y^2 E) &=& \frac{1}{4} (3 A^{\mu}C-C A^{\mu})\left(\frac{1}{4}A^2 +\frac{1}{2}\pa\cdot A\right) + \frac{1}{4} (5 A^2 \delta^{\mu}_{~\rho} - 4 A_{\rho} A^{\mu} +2 A^{\mu}A_{\rho})V^{\rho} ,\nn\\
\beta^{\mu}(-2 Y^{\rho}E Y_{\rho}) &=& \frac{1}{2}C\left[\left(\frac{1}{4}A^2 +\frac{1}{2}\pa\cdot A\right) A^{\mu} + (A^{*\mu} + \pa^{\mu}{\bar C})C\right]
\nn\\
&{}& -\frac{1}{2}(2A^{\rho} \delta^{\mu}_{~\nu} - A^{\mu}\delta^{\rho}_{~\nu}
- A_{\nu}\delta^{\mu\rho})\left[V^{\nu}A_{\rho} + 
\left(E^{\nu}_{~\rho} - \frac{A^2}{4}\delta^{\nu}_{~\rho}\right)C\right],\nn\\
\beta^{\mu}(EY^2) &=&  \frac{1}{4}V^{\mu} A^2 -\frac{1}{4}
\left(E^{\mu}_{~\rho}- \frac{A^2}{4}\delta^{\mu}_{~\rho}\right)(3 A^{\rho}C-C A^{\rho}) -(A^{*\mu} + \pa^{\mu}{\bar C})C^2.
\nn 
\eea
In \bref{alpha} and \bref{beta-mu}, 
\bea
D_{\rho} C &=& \pa_{\rho} C + [A_{\rho},~C], ~~~~~~~ 
F_{\rho\s}=[D_{\rho},~D_{\s}],\nn\\
R^{\mu\nu}_{\rho\s}&=& \biggl[ F_{\rho\s}\delta^{\mu\nu} -
\left(D_{\rho}A^{\mu}-\frac{1}{2}A^{\mu}A_{\rho}\right)\delta^{\nu}_{~\s}
-
\left(D_{\rho}A^{\nu}+\frac{1}{2}A_{\rho}A^{\nu}\right)\delta^{\mu}_{~\s}
\nn\\
&{}& + \frac{1}{2}A^2 \delta^{\mu}_{~\rho}\delta^{\nu}_{~\s}\biggr]
-\biggl[(\rho \leftrightarrow \s)\biggr].
\label{Rmunu}
\eea
We then obtain from \bref{matrix-ele}, \bref{nabla-E}, \bref{alpha},
\bref{beta-mu} and \bref{Rmunu}
\bea
{\rm tr}\left(-C\alpha + A_{\mu}\beta^{\mu}\right) &=&2~ {\rm tr}\biggl[C
\pa^2 \pa\cdot A - 2 (\pa C)\cdot A \pa\cdot A 
+ 2 (\pa^{\rho}C)A^{\s}\pa_{\rho}A_{\s}\nn\\
&{}& 
-2(\pa^{\rho}C)A^{\s}\pa_{\s}A_{\rho} +2(\pa^{\rho}C)A^{\s}A_{\rho}A_{\s}
\biggr].
\label{trfinal}
\eea
The sum of \bref{M^2-term2} and \bref{trfinal} yields \bref{mom-int1}.

%%%%%%%%%%%%%%%%%%%%%%%%%%%%%%%%%%%%%%%%%%%%%%%%%%%%%%%%%%%%%%%%%%%%%
%%%%%%%%%%%%%%%%%%%%%%%%%%%%%%%%%%%%%%%%%%%%%%%%%%%%%%%%%%%%%%%%%%%%%
%%%%%%%%%%%%%%%%%%%%%%%%%%%%%%%%%%%%%%%%%%%%%%%%%%%%%%%%%%%%%%%%%%%%%

\end{document}